**Thermal properties of graphene: Fundamentals and applications**

Eric Pop[1], Vikas Varshney[2], and Ajit K. Roy[3]

[1]University of Illinois at Urbana-Champaign ; epop@illinois.edu

[2]Vikas Varshney, Air Force Research Laboratory ; vikas.varshney@wpafb.af.mil

[3]Ajit K. Roy, Air Force Research Laboratory ; ajit.roy@wpafb.af.mil

Graphene is a two-dimensional (2D) material with over 100-fold anisotropy of heat flow between the in-plane and out-of-plane directions. High in-plane thermal conductivity is due to covalent $sp^2$ bonding between carbon atoms, whereas out-of-plane heat flow is limited by weak van der Waals coupling. Herein, we review the thermal properties of graphene, including its specific heat and thermal conductivity (from diffusive to ballistic limits) and the influence of substrates, defects, and other atomic modifications. We also highlight practical applications in which the thermal properties of graphene play a role. For instance, graphene transistors and interconnects benefit from the high in-plane thermal conductivity, up to a certain channel length. However, weak thermal coupling with substrates implies that interfaces and contacts remain significant dissipation bottlenecks. Heat flow in graphene or graphene composites could also be tunable through a variety of means, including phonon scattering by substrates, edges or interfaces. Ultimately, the unusual thermal properties of graphene stem from its 2D nature, forming a rich playground for new discoveries of heat flow physics and potentially leading to novel thermal management applications.

## Introduction

Graphene is a two-dimensional (2D) material, formed of a lattice of hexagonally arranged carbon atoms. Graphene is typically referred to as a single layer of graphite, although common references also exist to bilayer or trilayer graphene. (See the introductory article in this issue.) Most thermal properties of graphene are derived from those of graphite and bear the imprint of the highly anisotropic nature of this crystal.[1] For instance, the in-plane covalent $sp^2$ bonds between adjacent carbon atoms are among the strongest in nature (slightly stronger than the $sp^3$ bonds in diamond), with a bonding energy of approximately[2] 5.9 eV. By contrast, the adjacent graphene planes within a graphite crystal are linked by weak van der Waals interactions[2] (~50 meV) with a spacing[3] of $h \approx 3.35$ Å. **Figure 1**a displays the typical ABAB (also known as Bernal) stacking of graphene sheets within a graphite crystal.

The strong and anisotropic bonding and the low mass of the carbon atoms give graphene and related materials unique thermal properties. In this article we survey these unusual properties and their connection with the character of the underlying lattice vibrations. We examine both specific heat and thermal conductivity of graphene and related materials, and the conditions for achieving ballistic, scattering-free heat flow. We also investigate the role of atomistic lattice modifications and defects in tuning the thermal properties of graphene. Finally we explore the role of heat conduction in potential device applications and the possibility of architectures that allow control over the thermal anisotropy.

## Phonon dispersion of graphene

To understand the thermal properties of graphene, we must first inspect the lattice vibrational modes (phonons) of the material. The graphene unit cell, marked by dashed lines in Figure 1a, contains $N = 2$ carbon atoms. This leads to the formation of three acoustic (A) and $3N - 3 = 3$ optical (O) phonon modes, with the dispersions[4–7] shown in Figure 1b. The dispersion is the relationship between the phonon energy $E$ or frequency $\omega$ ($E = \hbar\omega$) and the phonon wave vector $q$. Longitudinal (L) modes correspond to atomic displacements along the wave propagation direction (compressive waves), while transverse (T) modes correspond to in-plane displacements perpendicular to the propagation direction (shear waves). In typical three dimensional (3D) solids transverse modes can have two equivalent polarizations, but the unique 2D nature of graphene allows out-of-plane atomic displacements, also known as flexural (Z) phonons.

At low $q$ near the center of the Brillouin zone, the frequencies of the transverse acoustic (TA) and longitudinal acoustic (LA) modes have linear dispersions,[8,9] $\omega_{TA} \approx v_{TA} q$ and $\omega_{LA} \approx v_{LA} q$. The group velocities $v_{TA} \approx 13.6$ km/s and $v_{LA} \approx 21.3$ km/s are four to six times higher than those in Si or Ge because of the strong in-plane $sp^2$ bonds of graphene and the small mass of carbon atoms.[8–11] In contrast, the flexural ZA modes have an approximately quadratic dispersion,[8,9] $\omega_{ZA} \approx \alpha q^2$ where $\alpha \approx 6.2 \times 10^{-7}$ m²/s. As we will discuss, the existence and modifications of these ZA modes are responsible for many of the unusual thermal properties of graphene.

**Specific heat of graphene and graphite**

The specific heat $C$ of a material represents the change in energy density $U$ when the temperature changes by one Kelvin, $C = dU/dT$, where $T$ is the absolute temperature. The specific heat and heat capacity are sometimes interchangeably used, with units of Joules per Kelvin per unit mass, or per volume or mole. The specific heat determines not only the thermal energy stored within a body but also how quickly the body cools or heats, that is, its thermal time constant $\tau \approx RCV$, where $R$ is the thermal resistance for heat dissipation (the inverse of conductance, $R = 1/G$) and $V$ is the volume of the body. Thermal time constants can be very short for nanoscale objects, on the order of 10 ns for nanoscale transistors,[12] 0.1 ns for a single graphene sheet or carbon nanotube (CNT),[13] and 1 ps for the relaxation of individual phonon modes.[14–16]

The specific heat of graphene has not been measured directly; thus, the short discussion here refers to experimental data available for graphite.[17–19] The specific heat is stored by the lattice vibrations (phonons) and the free conduction electrons of a material, $C = C_p + C_e$. However, phonons dominate the specific heat of graphene at all practical temperatures[19,20] (>1 K), and the phonon specific heat increases with temperature,[17–20] as shown in **Figure 2**. At very high temperatures[21] (approaching the in-plane Debye temperature[17,22] $\Theta_D \approx 2100$ K), the specific heat is nearly constant at $C_p = 3N_A k_B \approx 25$ J mol$^{-1}$ K$^{-1}$ $\approx 2.1$ J g$^{-1}$ K$^{-1}$, also known as the Dulong–Petit limit. Here $N_A$ is Avogadro's number and $k_B$ is the Boltzmann constant. This is the "classical" behavior of solids at high temperature when all six atomic degrees of motion (three translational and three vibrational) are excited and each carries $^1/_2 k_B T$ energy.

At room temperature, the specific heat of graphite is $C_p \approx 0.7$ J g$^{-1}$ K$^{-1}$, approximately one-third of the classical upper limit.[17,19] Interestingly, this value for graphite at room temperature is ~30% higher than that of diamond because of the higher density of states at low phonon frequencies given by the weak coupling between graphite layers.[17] A similar behavior is expected for an isolated graphene sheet at room temperature, when all of its flexural ZA modes should be thermally excited. However, it is possible that these modes could be partly suppressed or their dispersion altered when graphene is in strong contact with a substrate (thus lowering the specific heat), as suggested by experiments investigating epitaxial graphene on metals[23,24] and recent theoretical work on graphene on insulators.[25]

At low temperatures (**Figure 2** inset), the specific heat of a material scales as $C_p \sim T^{d/n}$ for a phonon dispersion $\omega \sim q^n$ in $d$ dimensions.[10,26] Thus, the low-temperature specific heat contains valuable information about both the dimensionality of a system and its phonon dispersion.[26] The $C_p$ of an isolated graphene sheet should be linear in $T$ at very low temperature when the quadratic ZA modes dominate, followed by a transition to $\sim T^2$ behavior from the linear LA and TA phonons[10,20,26] and eventually by a "flattening" to a constant as the high Debye temperature $\Theta_D$ is approached, in the classical limit (**Figure 2**). Indeed, numerical calculations using the complete phonon dispersion[10,26] reveal that, for a wide temperature range ($T < 50$ K), the $C_p$ of isolated graphene is linear in $T$ as shown in the **Figure 2** inset. By contrast, the specific heat of graphite rises as $\sim T^3$ at very low temperature (<10 K) because of the weak interlayer coupling[18] and then transitions to $\sim T^2$ behavior because of the in-plane linear phonons once the soft $c$-axis modes are fully occupied.[20] This behavior is consistent with graphite having both 2D and 3D features and is shown in the **Figure 2** inset. Calculations[19] and recent measurements[27] have also estimated the specific heat of the electronic gas in graphene at low temperature, finding values on the order of $C_e \approx 2.6$ μJ g$^{-1}$ K$^{-1}$ at 5 K (three orders of magnitude lower than the phonon specific heat, $C_p$, at this temperature; see **Figure 2**). The value of $C_e$ in graphene is lower than those in other 2D electron gases, opening up interesting opportunities for graphene as a sensitive bolometric detector.[27]

**Thermal conductivity of graphene: Intrinsic**

The thermal conductivity (κ) of a material relates the heat flux per unit area, $Q''$ (e.g., in W/m$^2$) to the temperature gradient, $Q'' = -\kappa \nabla T$. The sign in this relationship is negative, indicating that heat flows from high to low temperature. The thermal conductivity can be related to the specific heat by $\kappa \approx \sum C v \lambda$, where $v$ and λ are appropriately averaged phonon group velocity and mean free path, respectively.[28] This expression is commonly used under diffusive transport conditions, when sample dimensions are much greater than the mean free path ($L \gg \lambda$). (We discuss the ballistic heat-flow regime in a later section.) For the purposes of heat transport, the "thickness" of a graphene monolayer is typically assumed to be the graphite interlayer spacing,[3] $h \approx 3.35$ Å.

The in-plane thermal conductivity of graphene at room temperature is among the highest of any known material, about 2000–4000 W m$^{-1}$ K$^{-1}$ for freely suspended samples[29–31] (**Figures 3a-b**). The upper end of this range is achieved for isotopically purified samples (0.01% $^{13}$C instead of 1.1% natural abundance) with large grains,[31] whereas the lower end corresponds to isotopically mixed samples or those with smaller grain sizes. Naturally, any additional disorder or even residue from sample fabrication[32] will introduce more phonon scattering and lower these values further. For comparison, the thermal conductivity of natural diamond is ~2200 W m$^{-1}$ K$^{-1}$ at room temperature[33,34] (that of isotopically purified diamond is 50% higher, or ~3300 W m$^{-1}$ K$^{-1}$), and those of other related materials are plotted in Figures 3a-b. In particular, Figure 3b shows presently known *ranges* in thermal conductivity at room temperature, with the implication that all lower bounds could be further reduced in more disordered samples.

By contrast, heat flow in the cross-plane direction (along the *c* axis) of graphene and graphite is strongly limited by weak inter-plane van der Waals interactions. The thermal conductivity along the *c* axis of pyrolytic graphite is a mere ~6 W m$^{-1}$ K$^{-1}$ at room temperature,[1,30] as shown in Figure 3a. Heat flow perpendicular to a graphene sheet is also limited by weak van der Waals interactions with adjacent substrates, such as $SiO_2$. The relevant metric for heat flow across such interfaces is the thermal conductance per unit area, $G'' = Q''/\Delta T \approx 50$ MW m$^{-2}$ K$^{-1}$ at room temperature.[35–37] This is approximately equivalent to the thermal resistance of a ~25-nm layer of $SiO_2$[12] and could become a limiting dissipation bottleneck in highly scaled graphene devices and interconnects,[38] as discussed in a later section. Interestingly, the thermal resistance, $1/G''$, does not change significantly across few-layer graphene samples[36] (i.e., from one to 10 layers), indicating that the thermal resistance between graphene and its environment dominates that between individual graphene sheets. Indeed, the interlayer thermal conductance of bulk graphite is ~24 GW m$^{-2}$ K$^{-1}$ if the typical 3.35-Å spacing (Figure 1a) and the *c*-axis thermal conductivity are assumed.

**Thermal conductivity of graphene: Roles of edges and substrates**

Despite its high room-temperature value for freely suspended samples, the in-plane thermal conductivity of graphene decreases significantly when this 2D material is in contact with a substrate or confined into graphene nanoribbons (GNRs). This behavior is not unexpected, given that phonon propagation in an atomically thin graphene sheet is likely to be very sensitive to surface or edge perturbations. At room temperature, the thermal conductivity of graphene supported[39] by $SiO_2$ was measured as ~600 W m$^{-1}$ K$^{-1}$, that of $SiO_2$-encased graphene[40] was measured as ~160 W m$^{-1}$ K$^{-1}$, and that of supported GNRs[38] was estimated as ~80 W m$^{-1}$ K$^{-1}$ for ~20-nm-wide samples. The broader ranges of presently known values at room temperature are summarized in Figure 3b. Although differences could exist between these studies in terms of defects introduced during sample fabrication, for example, the results nevertheless suggest a clear decrease in thermal conductivity from that of isolated (freely suspended) graphene, consistent with theoretical predictions.[41–43]

For $SiO_2$-supported graphene, the decrease in thermal conductivity occurs as a result of the coupling and scattering of all graphene phonons with substrate vibrational modes,[16] the graphene ZA branch appearing to be most affected.[25,39] This decrease is also seen in Figure 3c, expressed as thermal conductance per cross-sectional area ($G/A$) which is a more appropriate measure when samples approach ballistic heat flow limits. For comparison, this figure also replots the thermal conductance of CNTs[44,45] and the theoretical upper limit of scattering-free ballistic transport ($G/A$)$_{ball}$ as calculated from the phonon dispersion.[8,11,74] (Also see the later section on ballistic transport.) Figure 3d illustrates the expected dependence of room-temperature thermal conductivity on sample length $L$ in a quasi-ballistic transport regime, as $L$ becomes comparable to the intrinsic phonon mean free path, $\lambda_0$. When graphene is confined into GNRs that are narrower than the intrinsic phonon mean free path ($W \leq \lambda_0$), phonon scattering with boundaries and edge roughness further reduces the thermal conductivity[42,43] compared to the case of suspended and $SiO_2$-supported graphene.

It is relevant to put such thermal properties of graphene in context. For comparison, the thermal conductivity of thin Si-on-insulator (SOI) films is also strongly reduced from the bulk Si value (~150 W m$^{-1}$ K$^{-1}$ at room temperature) to ~25 W m$^{-1}$ K$^{-1}$ in ~20-nm thin films as a result of surface scattering.[46] This value is further reduced to ~2 W m$^{-1}$ K$^{-1}$ in ~20-nm-diameter Si nanowires with rough surfaces.[47] At comparable linewidths, the thermal conductivity of Cu interconnects is on the order of ~100 W m$^{-1}$ K$^{-1}$ (a factor of four lower than that of bulk Cu) based on the Wiedemann–Franz law that relates thermal and electrical conductivity of metals.[48] In contrast, despite substrate or edge effects, graphene maintains a relatively high thermal conductivity in 2D monolayer films that are atomically thin ($h \approx 0.335$ nm), a size regime where no 3D materials can effectively conduct heat.

**Thermal modeling of graphene**

Given that thermal measurements of graphene are challenging because of its atomic thinness, modeling and simulation have played a key role in developing an understanding of graphene properties.[49] Existing methods for modeling thermal transport in graphene and GNRs include atomistic techniques such as molecular dynamics (MD),[16,25,50–56] non-equilibrium Green's functions (NEGF),[57–60] and Boltzmann transport equation (BTE) simulations.[9,39,41,43] The following discussion focuses on MD simulations, which have provided atomistic insights into graphene heat flow and have also predicted novel routes to tailor the thermal properties of nanostructured graphene materials.

*Insights from molecular dynamics*

MD is a deterministic approach for investigating properties of molecular systems that employs empirical interactions between atoms as a "force-field" and follows classical Newtonian dynamics.[61] **Figure 4**a schematically illustrates one of the two NEMD methodologies which is routinely used to investigate thermal transport in graphene or GNRs. In this methodology, atoms at both ends are kept fixed while near-end portions of few nm are treated as hot and cold regions (see **Figure 4**a). By imposing either constant heat flux or constant temperature boundary conditions in the hot and cold regions, a steady-state temperature gradient is introduced within the graphene sheet, which is then used to estimate the material thermal conductivity.

MD simulations have revealed how heat flow can be tuned or altered with respect to that of pristine graphene by introducing atomistic alterations of the lattice. Such alterations are achieved through vacancies or Stone–Wales defects,[55,62] grain boundaries,[63,64] strain,[65,66] chemical functionalization,[67] isotopic impurities ($^{13}$C)[52,53] or substitutional defects,[54] and edge roughness[50,51,53] or folding[60] in GNRs, as shown in **Figure 4**b. Alterations or defects can reduce the thermal conductivity of graphene by an order of magnitude or more below its intrinsic value, as summarized in **Table I**. Such a reduction in thermal conduction could be interesting for thermoelectric applications, if the high electronic conduction of graphene can be preserved.[68] For instance, in the limit of *zero* lattice contribution to thermal conductivity, the maximum thermoelectric figure of merit of a material is given only by its Seebeck coefficient $S$ and the Lorenz number $L_0$, $ZT = S^2/L_0 \sim 0.4$ assuming $S = 100$ µV/K.

Another interesting feature predicted by thermal MD simulations of graphene is that of thermal rectification. By analogy with electrical rectification in a *p–n* diode, a thermal rectifier would allow greater heat flux in one direction than another, that is, $Q_{BA} > Q_{AB}$ for the same temperature difference $\Delta T_{BA} = \Delta T_{AB}$ between its two terminals A and B.[12] Any type of spatial variability that introduces asymmetry in the phonon density of states of the hot and cold region has been identified as a key criterion necessary for thermal rectification. For graphene, such a feature has been identified via MD simulations by introducing either shape asymmetry within the nanostructure (such as a thickness-modulated GNR,[56] tapered-width GNR,[50,69] or Y-shaped GNR[70]) or mass asymmetry through substitution with [13]C isotopes.[71] In addition, a recent study has also suggested that asymmetry in thermal reservoirs is as essential as system asymmetry in achieving thermal rectification in any system.[72] No matter how it is achieved, such modulation of directional heat flux could provide novel functionality in future nanoelectronic devices such as thermal rectifiers, thermal transistors and thermal logic gates.

Nevertheless, the results of MD simulations should be interpreted in the proper context.[XX] The main strength of the MD approach is that it can be used to analyze the effects of atomistic changes on the thermal properties of a nanomaterial (**Figure 4** and **Table** I). However, MD is a semi-classical technique that overestimates the specific heat below the Debye temperature, $\Theta_D$. Graphene has a very high Debye temperature, $\Theta_D \approx 2100$ K, such that the specific heat at room temperature is only about one-third that of the classical Dulong–Petit limit (Figure 2). MD results are also sensitive to the choice of interatomic potential.[55,73] Thus, absolute values of thermal conductivity for graphene and GNR calculated by MD span a wide range (75–10,000 W m$^{-1}$ K$^{-1}$; see **Table** I) because of differences in interatomic potentials,[55,73] boundary conditions, and simulated system dimensions (often 10 nm or smaller). The effect of system dimensions is more challenging in graphene than in other materials because of the very large intrinsic phonon mean free path, $\lambda_0 \approx 600$ nm (see the next section). Thus, MD simulations should generally be interpreted based on the *relative* changes rather than the absolute values of the thermal properties they predict. Such changes are listed in the last column of Table I.

### Ballistic limit of graphene thermal conductivity

While the classical regime of large sample size ($L \gg \lambda_0$) suggests a constant thermal conductivity, $\kappa$, and a thermal conductance that scales inversely with length, $G = \kappa A/L$, a quantum treatment of small graphene devices ($L \ll \lambda_0$) reveals that the thermal conductance approaches a constant ($G_{ball}$), independent of length,[8,11,74] in ballistic, scattering-free transport. Thus, the relationship between conductivity and conductance imposes that the effective thermal conductivity of a ballistic sample must be proportional to its length as $\kappa_b = (G_{ball}/A)L$, where $A$ is the cross-sectional area, $A = Wh$. This is an important distinction also made between the *electrical* conductance, which reaches a constant (e.g., ~155 μS in single-walled CNTs with four quantum channels[75,76]), and the electrical conductivity and mobility, which appear to depend on the device length in the ballistic regime.[77,78]

The ballistic thermal conductance of graphene can be numerically calculated[8,11,74] from the phonon dispersion (Figure 1b) and is shown by the solid line in Figure 3c. This upper ballistic limit can also be approximated analytically[8] as $G_{ball}/A \approx 6 \times 10^5 \, T^{1.5} \, \text{W m}^{-2} \, \text{K}^{-5/2}$ for $T < 100$ K. The $\sim T^{1.5}$ dependence arises from the dominance of flexural ZA modes at low temperatures, with a specific heat $C \sim T$ and a phonon dispersion with $\omega \sim q^2$. A comparison with the experimental data available today in terms of conductance per unit area (symbols in Figure 3c) reveals that various measurements have all reached only a fraction of this ballistic limit. For instance, 10-μm-long graphene supported[39] on $SiO_2$ reached ~2%, and 2.8-μm long suspended graphene[31] samples reached ~25% of the theoretical ballistic thermal conductance limit at room temperature.

The transition of thermal conductivity from the ballistic ($L \ll \lambda_0$) to the diffusive ($L \gg \lambda_0$) heat flow regime can be approximated through a Landauer-like approach[28,79] as $\kappa(L) \approx G_{ball}/A[1/L + 2/(\pi\lambda)]^{-1}$, where the factor of $\pi/2$ accounts for angle averaging[80] in 2D to obtain the backscattering length responsible for the thermal resistance. Fitting this simple expression to the experimental data in Figure 3d reveals phonon mean free paths at room temperature of $\lambda_0 \approx 600$ nm in suspended graphene (also known as the intrinsic mean free path), $\lambda \approx 100$ nm in graphene supported on $SiO_2$, and $\lambda \approx 20$ nm in GNRs (of width ~20 nm) supported on $SiO_2$. These are some of the key length scales needed for understanding graphene thermal properties in nanometer-size devices. The ballistic upper limit of thermal conductivity in a graphene sample of length $L \approx 100$ nm can now be estimated as $\kappa_b \approx 350$ W m$^{-1}$ K$^{-1}$ at room temperature. In addition, suspended graphene should attain >80% of the ballistic heat flow limit in samples shorter than $L < 235$ nm, whereas graphene supported on $SiO_2$ reaches this level at $L < 40$ nm, well within the means of modern nanofabrication.

**Thermal properties for applications**

*Devices and interconnects*

In the context of nanoscale devices and interconnects, graphene is often thought to hold advantages over other materials because of its higher thermal conductivity. Thus, high thermal conductivity could superficially suggest very good heat sinking and low temperature rise during device operation. However, under high-field and high-temperature (i.e., typical circuit) operating conditions, significant dissipation and temperature rise can nevertheless occur in graphene devices,[38,81] as shown in **Figure 5**.

Self-heating of graphene devices and interconnects at high field begins through the emission of optical phonons (OPs),[82–84] similarly to the case of CNTs. OPs are strongly emitted at applied voltages comparable to or greater than their energy (~0.16 eV; see **Figure 1b**), although smaller biases can also be sufficient due to the long Fermi tail of the electron (or hole) distribution. OPs decay on time scales of ~1 ps into lower-energy acoustic phonons (APs).[14,85] However, given their comparatively large specific heat, the AP temperature lags behind that of the electrons and OPs by ~1–10 ns after a voltage pulse is applied. (This delay also depends on the thermal resistance between the device and the surrounding environment.[12])

The pathway of heat dissipation to the environment heat sink becomes key in determining the temperature rise once steady state is reached and thus, ultimately, the reliability of graphene devices. In other words, despite (or perhaps because of) the excellent intrinsic thermal properties of graphene, dissipation from graphene devices is often limited by their interfaces, contacts, and surrounding materials, which are often thermal insulators such as $SiO_2$. To illustrate this point, **Figure 5**a shows temperature profiles recorded by infrared (IR) thermal imaging[81] along a graphene device on $SiO_2$ under a constant source–drain bias ($V_{DS}$ = -12 V) as the gate voltage ($V_{GS}$) is varied from –5 V to 4 V. The complex temperature profile occurs because the carrier density and, thus, the electric field are not constant along the device at high bias. Consequently, the temperature hot spot marks the location of maximum electric field and minimum carrier concentration.[81]

A schematic of dissipation in a graphene device is shown in **Figure 5**b, where heat flow can occur either into the substrate or to the metal contacts.[38,86] The length scale for lateral heat flow to the contacts is the thermal healing length $L_H \approx (\kappa W h/g)^{1/2}$, where $W$ is the device width, $g$ is the thermal conductance to the substrate per unit length,[38] and other symbols are as previously defined. The total thermal conductance $g$ includes the contribution from the graphene-substrate interface, and that of any underlying layers (e.g. $SiO_2$ and Si in Figure 5b). For typical supporting oxide thicknesses ($t_{ox} \approx$ 90–300 nm) and interfacial thermal conductance $G''$, $L_H \approx 0.1$ μm.

Numerical calculations suggest that only devices shorter than ~$3L_H \approx 0.3$ μm benefit from substantial cooling through the metal contacts.[38] For "long" devices ($L \gg 3L_H$), the dissipation occurs almost entirely through the graphene–substrate interface (of thermal resistance $1/G''$) and through the underlying substrate (e.g., $SiO_2$/Si, BN/Si, SiC). For "narrow" devices ($W < t_{ox}$) such as GNRs, a substantial amount of lateral heat spreading into the underlying oxide can also play a role[38], as illustrated in Figure 5c. Finally, for devices that are both "long and wide" ($L, W \gg L_H, t_{ox}$), the total thermal resistance can be estimated simply as[83] $R_{th} \approx 1/(G''A) + t_{ox}/(\kappa_{ox}A) + 1/(2\kappa_{Si}A^{1/2})$, where $\kappa_{ox}$ and $\kappa_{Si}$ are the thermal conductivities of $SiO_2$ and Si, respectively, $A = LW$ is the device area, and other variables are as defined in Figure 5. The final term approximates the spreading thermal resistance into the Si substrate, which is assumed to be much thicker than both $t_{ox}$ and the graphene device dimensions. We note that improved heat sinking can be obtained by placing devices on substrates with a thinner supporting insulator or higher thermal conductivity, as long as the graphene–substrate interface is not the limiting factor.[35–37]

Recent work has also suggested that graphene devices might benefit from thermoelectric (Peltier) cooling at the metal contacts,[86] where a substantial difference in Seebeck coefficient exists. However, it is important to realize that, because of the one-dimensional (1D) nature of current flow, Peltier effects of opposite sign will occur at the two contacts, such that one cools as the other heats. Thus, additional contact engineering must be done to adjust the overall device temperature, for example, using asymmetric contacts, either from the point of view of either geometry (one larger contact to sink heat) or materials (two contacts with different Seebeck coefficients).

### 3D architectures

As summarized earlier, because of its 2D nature, graphene has very high anisotropy of its thermal properties between the in-plane and out-of-plane directions. Whereas the in-plane thermal conductivity is excellent (>1000 W m$^{-1}$ K$^{-1}$), the out-of-plane thermal coupling is limited by weak van der Waals interactions and could become a thermal dissipation bottleneck. To overcome this in practice, 3D architectures could incorporate CNT–pillared graphene network (PGN) structures,[87] interconnected CNT truss-like structures,[88] and networked graphene flakes.[89] These 3D architectures (**Figure 6**) are envisioned as a new generation of nanomaterials with tunable thermomechanical functionality, leveraging the best aspects of both graphene and CNTs. Such structures could have numerous applications, enabling efficient electrodes for fuel cells,[90] nanoporous structures with very high surface area for hydrogen storage,[87] supercapacitors,[91] and tailored multidimensional thermal transport materials.

From a thermal transport perspective, recent modeling studies suggest that the lateral CNT separation, called the interjunction distance (IJD), and the interlayer distance (ILD) between graphene sheets play a critical role in determining the thermal transport properties in these 3D architectures.[92,93] When the lateral CNT separation, IJD, is on the order of tens of nanometers, the ballistic nature of heat propagation (because of the large phonon mean free path in graphene and CNTs) causes phonon scattering to occur primarily at the CNT-graphene junction nodes. These junctions, in turn, will govern the thermal conductivity of such architectures. Furthermore, as the carbon atoms and $sp^2$ bonds of CNTs and graphene are the same, the phonon spectra are similar and junctions have very low interface thermal resistance. Hence, the thermal transport in different directions could be manipulated by tailoring the IJDs and ILDs.

For instance, the predicted interface thermal conductance at a junction[63,64] (~10 GW m$^{-2}$ K$^{-1}$) is comparable to that between graphite layers (~24 GW m$^{-2}$ K$^{-1}$) and over two orders of magnitude higher than the graphene thermal coupling with a substrate (~50 MW m$^{-2}$ K$^{-1}$ at room temperature[35–37]). This suggests that very dense packing of long CNTs (i.e., small IJD, large ILD) could significantly increase the out-of-plane thermal conductivity of the PGN architecture, by reducing the number of interfaces and replacing them with CNTs.[91] On the other extreme, using short but widely spaced CNTs in the PGN structure would substantially reduce thermal conduction in the out-of-plane direction[91] (due to the small ILD, higher interface density, and low CNT areal density), thus possibly opening several routes for thermoelectric applications where extremely low thermal conductivity is desired. Over the past few years, multiple research groups have successfully synthesized CNT pillared-graphene architectures, and different property characterizations are underway.[91,94–96]

## Summary

In summary, the unusual thermal properties of graphene include very high in-plane thermal conductivity (strongly affected by interfacial interactions, atomic defects, and edges) and relatively low out-of-plane thermal conductance. The specific heat of graphene is dominated by phonons and is slightly higher than that of graphite and diamond below room temperature. The in-plane thermal conductance $G$ of graphene can reach a significant fraction of the theoretical ballistic limit in sub-micron samples, owing to the large phonon mean free path ($\lambda \approx 100$–$600$ nm in supported and suspended samples, respectively). Nevertheless this behavior leads to an apparent dependence of thermal conductivity $\kappa$ on sample length, similar to the behavior of mobility in quasi-ballistic electronic devices.

In the context of integrated electronics, heat dissipation from graphene devices and interconnects is primarily limited by their environment and the relatively weak van der Waals interfaces of graphene. In the context of graphene composites and 3D architectures, simulation results have suggested that the thermal properties could be highly tunable. Such tunability raises the interesting prospects of both ultra-high thermal conductivity for heat sinking applications, and of ultra-low thermal conductivity for thermoelectric applications.

## Acknowledgments

E.P. acknowledges assistance from Zuanyi Li, Andrey Serov, Pierre Martin, Ning Wang, Albert Liao, Changwook Jeong, and Mark Lundstrom, as well as funding from the AFOSR Young Investigator Program (YIP) award, the ARO Presidential Early Career (PECASE) award, and the Nanotechnology Research Initiative (NRI). V.V. and A.K.R acknowledge Dr. Byung-Lip (Les) Lee (AFOSR Task 2302BR7P) and Dr. Joycelyn Harrison (AFOSR Task 2306CR9P) for funding from AFOSR.

**References**

1. H.O. Pierson, *Handbook of Carbon, Graphite, Diamond and Fullerenes: Properties, Processing and Applications* (Noyes Publications, Park Ridge, NJ, 1993).

2. M.C. Schabel, J.L. Martins, *Phys. Rev. B* **46**, 7185 (1992).

3. D.W. Bullett, *J. Phys. C: Solid State Phys.* **8**, 2707 (1975).

4. R. Saito, G. Dresselhaus, M.S. Dresselhaus, *Physical Properties of Carbon Nanotubes* (World Scientific, Singapore, 1998).

5. M. Mohr, J. Maultzsch, E. Dobardžić, S. Reich, I. Milošević, M. Damnjanović, A. Bosak, M. Krisch, C. Thomsen*, Phys. Rev. B* **76**, 035439 (2007).

6. C. Oshima, T. Aizawa, R. Souda, Y. Ishizawa, Y. Sumiyoshi, *Solid State Commun.* **65**, 1601 (1988).

7. L. Wirtz, A. Rubio, *Solid State Commun.* **131**, 141 (2004).

8. N. Mingo, D.A. Broido, *Phys. Rev. Lett.* **95**, 096105 (2005).

9. D.L. Nika, E.P. Pokatilov, A.S. Askerov, A.A. Balandin, *Phys. Rev. B* **79**, 155413 (2009).

10. V.N. Popov, *Phys. Rev. B* **66**, 153408 (2002).

11. E. Muñoz, J. Lu, B.I. Yakobson, *Nano Lett.* **10**, 1652 (2010).

12. E. Pop, *Nano Res.* **3**, 147 (2010).

13. Z.-Y. Ong, E. Pop, *J. Appl. Phys.* **108**, 103502 (2010).

14. Z.-Y. Ong, E. Pop, J. Shiomi, *Phys. Rev. B* **84**, 165418 (2011).

15. K. Kang, D. Abdula, D.G. Cahill, M. Shim, *Phys. Rev. B* **81**, 165405 (2010).

16. B. Qiu, X. Ruan, *Appl. Phys. Lett.* **100**, 193101 (2012).

17. T. Tohei, A. Kuwabara, F. Oba, I. Tanaka, *Phys. Rev. B* **73**, 064304 (2006).

18. R. Nicklow, N. Wakabayashi, H.G. Smith, *Phys. Rev. B* **5**, 4951 (1972).

19. T. Nihira, T. Iwata, *Phys. Rev. B* **68**, 134305 (2003).

20. L.X. Benedict, S.G. Louie, M.L. Cohen, *Solid State Commun.* **100**, 177 (1996).

21. L.E. Fried, W.M. Howard, *Phys. Rev. B* **61**, 8734 (2000).

22. V.K. Tewary, B. Yang, *Phys. Rev. B* **79**, 125416 (2009).

23. T. Aizawa, R. Souda, Y. Ishizawa, H. Hirano, T. Yamada, K.-i. Tanaka, C. Oshima, *Surf. Sci.* **237**, 194 (1990).

24. A.M. Shikin, D. Farías, K.H. Rieder, *Europhys. Lett.* **44**, 44 (1998).

25. Z.Y. Ong, E. Pop, *Phys. Rev. B* **84**, 075471 (2011).

26. J. Hone, *Top. Appl. Phys.* **80**, 273 (2001).

27. K.C. Fong, K.C. Schwab, "Ultra-sensitive and Wide Bandwidth Thermal Measurements of Graphene at Low Temperatures," in press, *Phys. Rev. X* (available at http://arxiv.org/abs/1202.5737).

28. C. Jeong, S. Datta, M. Lundstrom, *J. Appl. Phys.* **109**, 073718 (2011).

29. S. Chen, A.L. Moore, W. Cai, J.W. Suk, J. An, C. Mishra, C. Amos, C.W. Magnuson, J. Kang, L. Shi, R.S. Ruoff, *ACS Nano* **5**, 321 (2010).

30. A.A. Balandin, *Nat. Mater.* **10**, 569 (2011).

31. S. Chen, Q. Wu, C. Mishra, J. Kang, H. Zhang, K. Cho, W. Cai, A.A. Balandin, R.S. Ruoff, *Nat Mater* **11**, 203 (2012).

32. M.T. Pettes, I. Jo, Z. Yao, L. Shi, *Nano Lett.* **11**, 1195 (2011).

33. T.R. Anthony, W.F. Banholzer, J.F. Fleischer, L.H. Wei, P.K. Kuo, R.L. Thomas, R.W. Pryor, *Phys. Rev. B* **42**, 1104 (1990).

34. R. Berman, *Phys. Rev. B* **45**, 5726 (1992).

35. Z. Chen, W. Jang, W. Bao, C.N. Lau, C. Dames, *Appl. Phys. Lett.* **95**, 161910 (2009).

36. Y.K. Koh, M.-H. Bae, D.G. Cahill, E. Pop, *Nano Lett.* **10**, 4363 (2010).

37. K.F. Mak, C.H. Lui, T.F. Heinz, *Appl. Phys. Lett.* **97**, 221904 (2010).

38. A.D. Liao, J.Z. Wu, X.R. Wang, K. Tahy, D. Jena, H.J. Dai, E. Pop, *Phys. Rev. Letters* **106**, 256801 (2011).

39. J.H. Seol, I. Jo, A.L. Moore, L. Lindsay, Z.H. Aitken, M.T. Pettes, X.S. Li, Z. Yao, R. Huang, D. Broido, N. Mingo, R.S. Ruoff, L. Shi, *Science* **328**, 213 (2010).

40. W. Jang, Z. Chen, W. Bao, C.N. Lau, C. Dames, *Nano Lett.* **10**, 3909 (2010).

41. L. Lindsay, D.A. Broido, N. Mingo, *Phys. Rev. B* **82**, 115427 (2010).

42. J. Haskins, A. Kınacı, C. Sevik, H.l. Sevinçli, G. Cuniberti, T. Çağın, *ACS Nano* **5**, 3779 (2011).

43. Z. Aksamija, I. Knezevic, *Appl. Phys. Lett.* **98**, 141919 (2011).

44. E. Pop, D. Mann, Q. Wang, K.E. Goodson, H.J. Dai, *Nano Lett.* **6**, 96 (2006).

45. P. Kim, L. Shi, A. Majumdar, P.L. McEuen, *Phys. Rev. Lett.* **87**, 215502 (2001).

46. W. Liu, M. Asheghi, *J. Appl. Phys.* **98**, 123523 (2005).

47. R. Chen, A.I. Hochbaum, P. Murphy, J. Moore, P. Yang, A. Majumdar, *Phys. Rev. Lett.* **101**, 105501 (2008).

48. W. Steinhögl, G. Schindler, G. Steinlesberger, M. Traving, M. Engelhardt, *J. Appl. Phys.* **97**, 023706 (2005).

49. P.G. Klemens, D.F. Pedraza, *Carbon* **32**, 735 (1994).

50. J.N. Hu, X.L. Ruan, Y.P. Chen, *Nano Lett.* **9**, 2730 (2009).

51. W.J. Evans, L. Hu, P. Keblinski, *Appl. Phys. Lett.* **96**, 203112 (2010).

52. H.J. Zhang, G. Lee, A.F. Fonseca, T.L. Borders, K. Cho, *J. Nanomater.* **2010**, 537657 (2010).

53. J.N. Hu, S. Schiffli, A. Vallabhaneni, X.L. Ruan, Y.P. Chen, *Appl. Phys. Lett.* **97**, 133107 (2010).

54. B. Mortazavi, A. Rajabpour, S. Ahzi, Y. Remond, S.M.V. Allaei, *Solid State Commun.* **152**, 261 (2012).

55. H.J. Zhang, G. Lee, K. Cho, *Phys. Rev. B* **84**, 115460 (2011).

56. W.-R. Zhong, W.-H. Huang, X.-R. Deng, B.-Q. Ai, *Appl. Phys. Lett.* **99**, 193104 (2011).

57. Y. Xu, X.B. Chen, J.S. Wang, B.L. Gu, W.H. Duan, *Phys. Rev. B* **81**, 195425 (2010).

58. Z. Huang, T.S. Fisher, J.Y. Murthy, *J. Appl. Phys.* **108**, 094319 (2010).

59. J.W. Jiang, B.S. Wang, J.S. Wang, *Appl. Phys. Lett.* **98**, 113114 (2011).

60. N. Yang, X. Ni, J.-W. Jiang, B. Li, *Appl. Phys. Lett.* **100**, 093107 (2012).

61. D. Frenkel, B. Smit, *Understanding Molecular Simulation: From Algorithms to Applications* (Academic Press, New York, ed. 2, 2002).

62. F. Hao, D.N. Fang, Z.P. Xu, *Appl. Phys. Lett.* **99**, 041901 (2011).

63. A. Bagri, S.P. Kim, R.S. Ruoff, V.B. Shenoy, *Nano Lett.* **11**, 3917 (2011).

64. A. Cao, J. Qu, *J. Appl. Phys.* **111**, 053529 (2012).

65. X. Li, K. Maute, M.L. Dunn, R. Yang, *Phys. Rev. B* **81**, 245318 (2010).

66. N. Wei, L. Xu, H.-Q. Wang, J.-C. Zheng, *Nanotechnology* **22**, 105705 (2011).

67. S.-K. Chien, Y.-T. Yang, C.O.-K. Chen, *Carbon* **50**, 421 (2012).

68. H. Sevinçli, G. Cuniberti, *Phys. Rev. B* **81**, 113401 (2010).

69. N. Yang, G. Zhang, B.W. Li, *Appl. Phys. Lett.* **95**, 033107 (2009).

70. G. Zhang, H.S. Zhang, *Nanoscale* **3**, 4604 (2011).

71. Q.-X. Pei, Y.-W. Zhang, Z.-D. Sha, V.B. Shenoy, *Appl. Phys. Lett.* **100**, 101901 (2012).

72. J. Lee, V. Varshney, A.K. Roy, J.B. Ferguson, B.L. Farmer, *Nano Letters* **12**, 3491 (2012).

73. L. Lindsay, D.A. Broido, *Phys. Rev. B* **81**, 205441 (2010).

74. K. Saito, J. Nakamura, A. Natori, *Phys. Rev. B* **76**, 115409 (2007).

75. A. Javey, J. Guo, M. Paulsson, Q. Wang, D. Mann, M. Lundstrom, H. Dai, *Phys. Rev. Lett.* **92**, 106804 (2004).

76. J.-Y. Park, S. Rosenblatt, Y. Yaish, V. Sazonova, H. Üstünel, S. Braig, T.A. Arias, P.W. Brouwer, P.L. McEuen, *Nano Lett.* **4**, 517 (2004).

77. M.S. Shur, *IEEE Electron Device Lett.* **23**, 511 (2002).

78. J. Wang, M. Lundstrom, *IEEE Trans. Electron Devices* **50**, 1604 (2003).

79. R. Prasher, *Phys. Rev. B* **77**, 075424 (2008).

80. C. Jeong, R. Kim, M. Luisier, S. Datta, M. Lundstrom, *J. Appl. Phys.* **107**, 023707 (2010).

81. M.-H. Bae, S. Islam, V.E. Dorgan, E. Pop, *ACS Nano* **5**, 7936 (2011).

82. A. Barreiro, M. Lazzeri, J. Moser, F. Mauri, A. Bachtold, *Phys. Rev. Lett.* **103**, 076601 (2009).

83. V.E. Dorgan, M.H. Bae, E. Pop, *Appl. Phys. Lett.* **97**, 082112 (2010).

84. V. Perebeinos, P. Avouris, *Phys. Rev. B* **81**, 195442 (2010).

85. K. Kang, D. Abdula, D.G. Cahill, M. Shim, *Phys. Rev. B* **81**, 165405 (2010).

86. K.L. Grosse, M.H. Bae, F.F. Lian, E. Pop, W.P. King, *Nat. Nanotechnol.* **6**, 287 (2011).

87. G.K. Dimitrakakis, E. Tylianakis, G.E. Froudakis, *Nano Lett.* **8**, 3166 (2008).

88. E. Tylianakis, G.K. Dimitrakakis, S. Melchor, J.A. Dobado, G.E. Froudakis, *Chem. Commun.* **47**, 2303 (2011).

89. Z.P. Chen, W.C. Ren, L.B. Gao, B.L. Liu, S.F. Pei, H.M. Cheng, *Nat. Mater.* **10**, 424 (2011).

90. W. Zhang, P. Sherrell, A.I. Minett, J.M. Razal, J. Chen, *Energy Environ. Sci.* **3**, 1286 (2010).

91. F. Du, D.S. Yu, L.M. Dai, S. Ganguli, V. Varshney, A.K. Roy, *Chem. Mater.* **23**, 4810 (2011).

92. V. Varshney, S.S. Patnaik, A.K. Roy, G. Froudakis, B.L. Farmer, *ACS Nano* **4**, 1153 (2010).

93. V. Varshney, A.K. Roy, G. Froudakis, B.L. Farmer, *Nanoscale* **3**, 3679 (2011).

94. R.K. Paul, M. Ghazinejad, M. Penchev, J.A. Lin, M. Ozkan, C.S. Ozkan, *Small* **6**, 2309 (2010).

95. L.L. Zhang, Z.G. Xiong, X.S. Zhao, *ACS Nano* **4**, 7030 (2010).

96. K.H. Yu, G.H. Lu, Z. Bo, S. Mao, J.H. Chen, *J. Phys. Chem. Lett.* **2**, 1556 (2011).

## Figure Captions

Figure 1. (a) Schematic of the atomic arrangement in graphene sheets. Dashed lines in the bottom sheet represent the outline of the unit cell. The areal density of carbon atoms in graphene is $3.82 \times 10^{15}$ cm$^{-2}$. (b) Graphene phonon dispersion obtained from the force constant method[4] along the $\Gamma$ to M crystallographic direction.[5,6] Note the presence of linear in-plane acoustic modes (longitudinal acoustic, LA; transverse acoustic, TA), as well as flexural out-of-plane acoustic (ZA) modes with a quadratic dispersion. The latter are responsible for many of the unusual thermal properties of graphene. Graphene has a much higher sound velocity and optical phonon (OP) energy than most materials; by comparison, OPs have energies of ~0.035 eV in Ge and GaAs and ~0.06 eV in Si. LO, longitudinal optical; TO, transverse optical; ZO, out-of-plane optical.

Figure 2. Specific heat of graphene, graphite, and diamond, all dominated by phonons at temperatures above ~1 K. Lines show numerical calculations,[10,17,26] symbols represent experimental data.[19,21] The inset indicates that the low-temperature specific heat of an *isolated* graphene sheet is expected to be higher than that of graphite due to the contribution of low-frequency ZA phonons (also see Figure 1b). Above ~100 K the specific heat of graphene and graphite should be identical. The inset makes use of different units to illustrate a common occurrence in practice (e.g. J mol$^{-1}$ K$^{-1}$, or J g$^{-1}$ K$^{-1}$, or J cm$^{-3}$ K$^{-1}$), but conversion is easily achieved by dividing and/or multiplying with the atomic mass of carbon ($A = 12.01$ g/mol) and the density of graphite ($\rho \approx 2.25$ g/cm$^3$).

Figure 3. (a) Thermal conductivity $\kappa$ as a function of temperature for representative data of suspended graphene ($\bigcirc$),[31] SiO$_2$-supported graphene ($\blacksquare$),[39] ~20-nm-wide graphene nanoribbons (GNRs, $\blacklozenge$),[38] suspended single-walled CNTs (SWCNTs, $+$),[44] multi-walled CNTs (MWCNTs, $\bullet$),[45] type IIa diamond ($\diamond$),[ZZ] graphite in-plane ($\triangleright$)[ZZ] and out-of-plane ($\triangle$).[ZZ] Additional data for graphene and related materials are also summarized in Refs. 30 and XX. (b) Room temperature ranges of thermal conductivity data $\kappa$ for diamond,[33] graphite (in-plane),[30] carbon nanotubes (CNTs),[30] suspended graphene,[30,31] SiO$_2$-supported graphene,[39] SiO$_2$-encased graphene,[40] and GNRs.[38] (c) Thermal conductance $G$ per cross-sectional area $A$ for graphene and related materials (symbols), compared to the theoretical ballistic limit, $G_{ball}/A$ (solid line).[8,11,74] (d) Expected scaling of thermal conductivity $\kappa$ with sample length $L$ in the quasi-ballistic regime, at $T \approx 300$ K. The solid line is the ballistic limit, $\kappa_{ball} = (G_{ball}/A)L$, and dashed lines represent $\kappa$ estimated with phonon mean free paths as labeled (see text), chosen to match experimental data for suspended graphene,[31] supported graphene,[39] and GNRs;[38] symbols are consistent with panels (a) and (c).

Figure 4. (a) Schematic of non-equilibrium molecular dynamics (NEMD) methodology for examining thermal transport in a graphene nanoribbon (GNR). (b) GNR showing different types of defects (vacancies, grain boundaries, Stone–Wales defects, substitutional and functionalization defects, and wrinkles or folds),

that have a profound effect in tuning thermal transport in graphene. Also see Table I.

Figure 5. (a) Infrared (IR) imaging of temperature in a functioning graphene field-effect transistor (GFET) with a drain bias $V_{DS} = -12$V and varying gate bias.[81] The device is back-gated, allowing IR imaging from the top. The hot spot marks the location of lowest carrier density (which changes with voltage bias) and highest electric field. (b) Longitudinal cross section of a graphene device or interconnect showing heat dissipation pathways (red arrows) and temperature profile $T(x)$. The device, of length $L$ and width $W$, is supported by an insulator (e.g. $SiO_2$) of thickness $t_{ox}$ on a Si substrate of thickness $t_{Si}$. The bottom of the substrate and the Pd contacts are assumed to be at temperature $T_0$. Significant heat can flow to the contacts within a distance of the thermal healing length $L_H$, reducing the temperature of devices shorter than ~$3L_H$, or $\leq$ ~0.3 μm. (c) Transverse cross-section showing heat dissipation from a narrow GNR ($W \ll t_{ox}$), which benefits from lateral heat spreading into the substrate and can carry peak current densities (~$10^9$ A/cm$^2$) higher than wide GNRs.[38]

Figure 6. Schematic of a three-dimensional (3D) nanoarchitecture that combines carbon nanotube pillars and graphene sheets to achieve tunable cross-plane thermal transport. For instance, reducing the interjunction distance (IJD) and increasing the interlayer distance (ILD) could mitigate the weak interlayer thermal coupling of a graphene stack for higher cross-plane thermal conductivity. Conversely, longer IJD and shorter ILD could lower cross-plane thermal conductivity, leading to thermal insulator or thermoelectric applications.

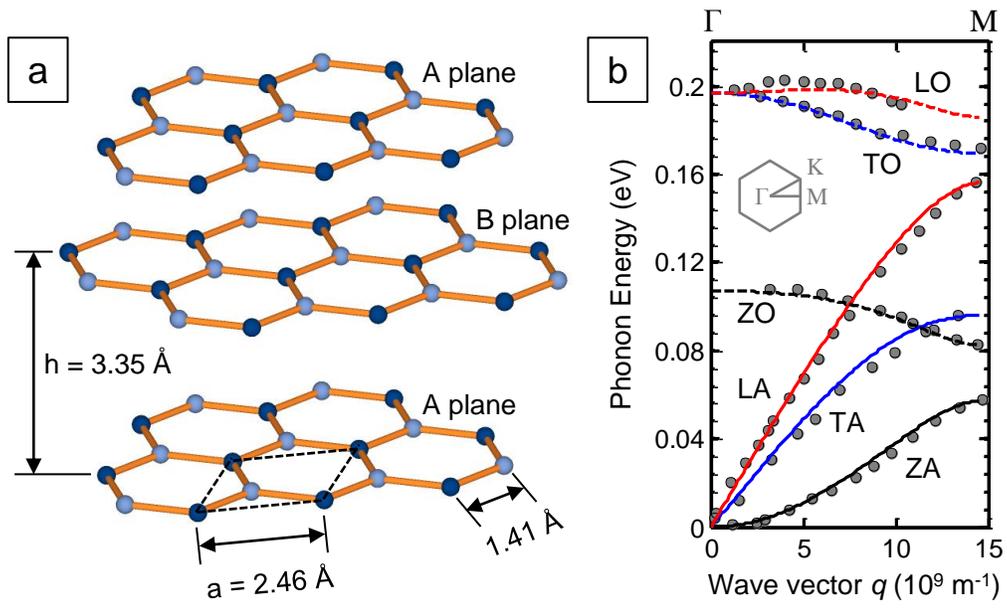

**Figure 1**

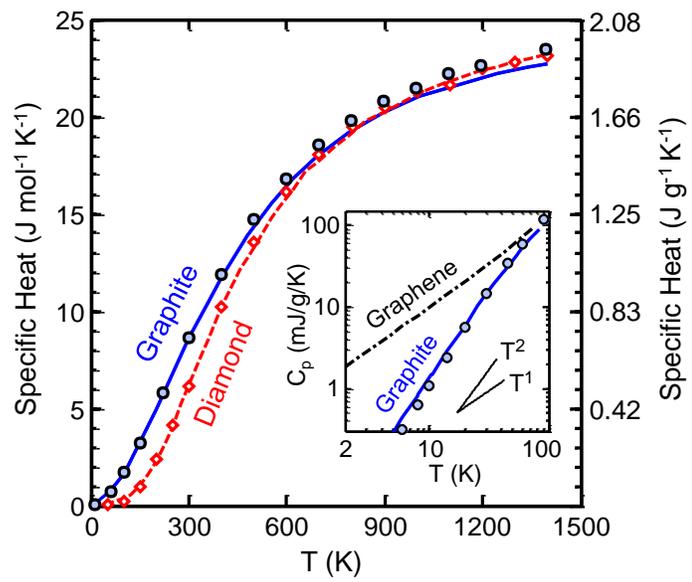

**Figure 2**

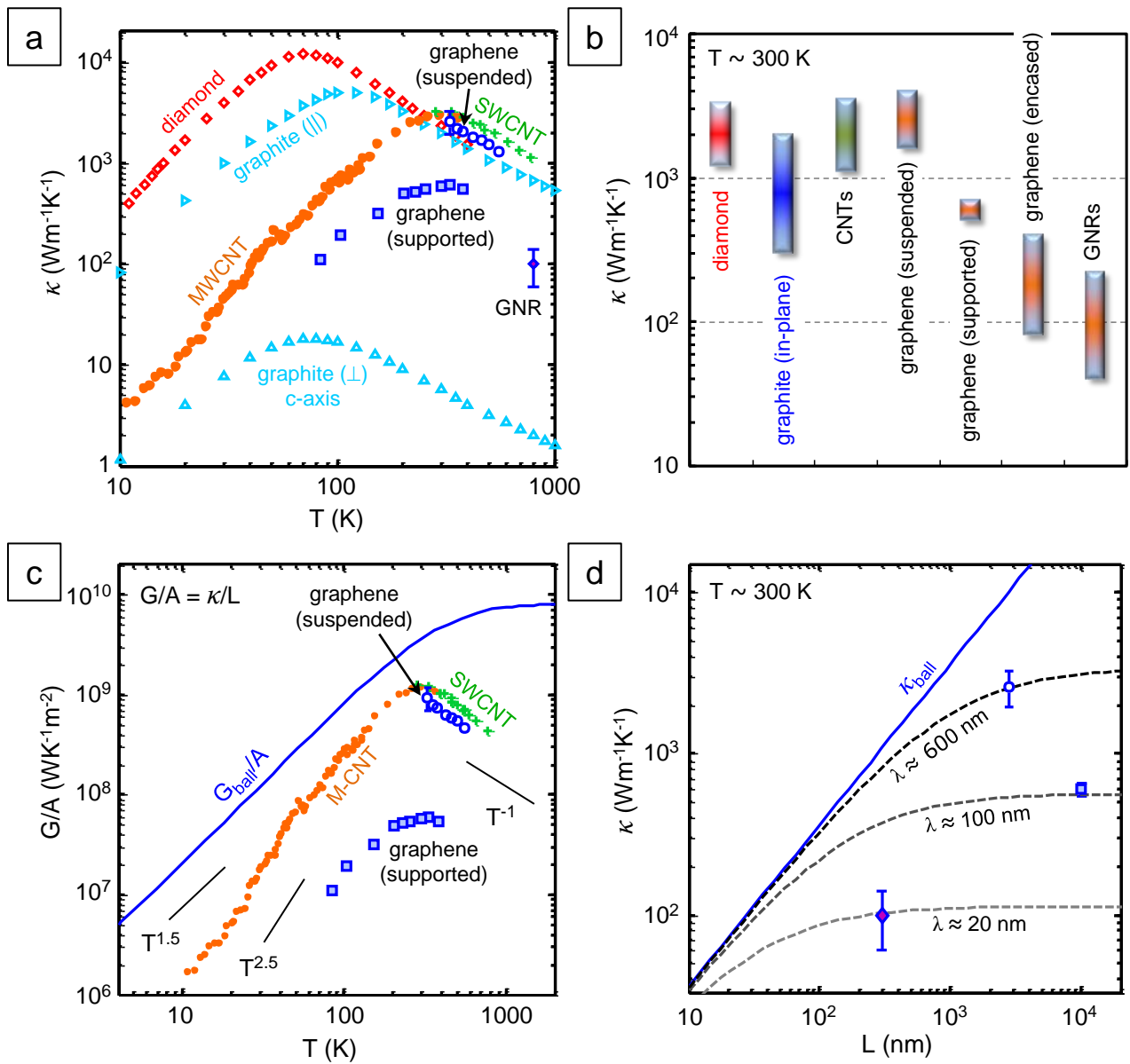

**Figure 3**

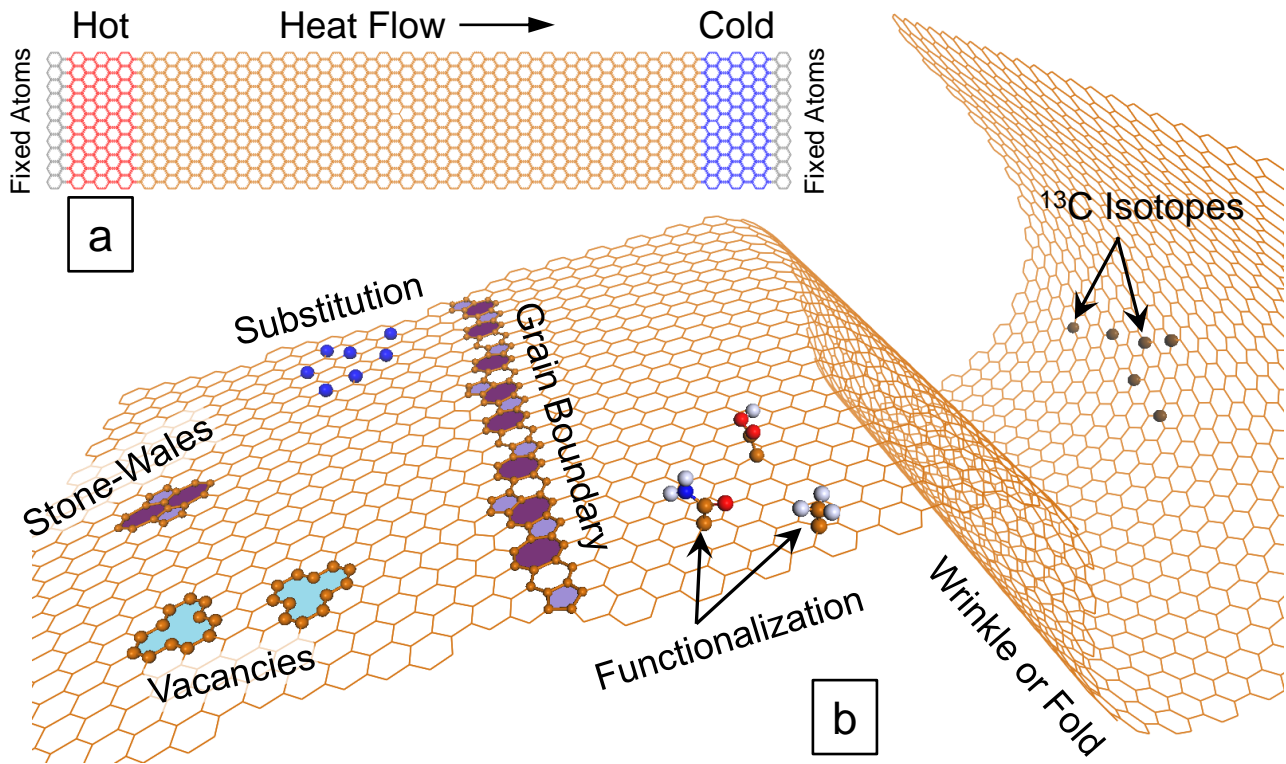

**Figure 4**

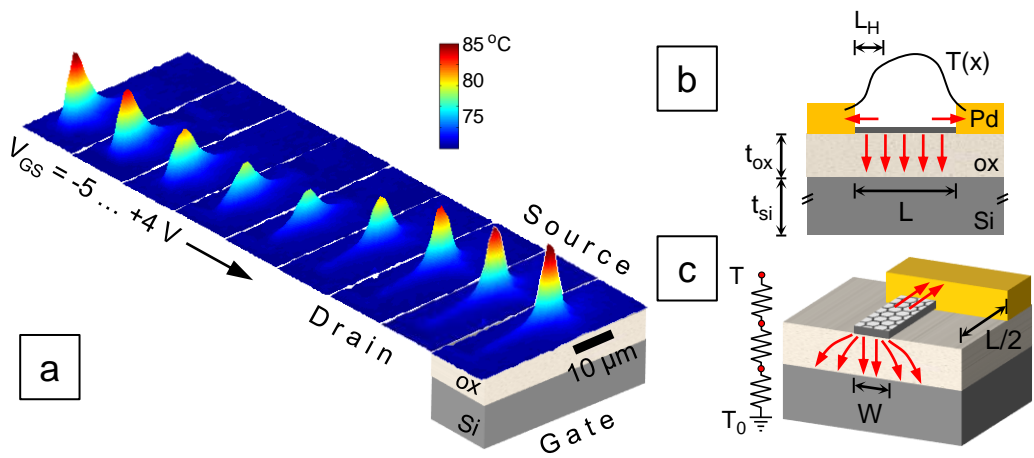

**Figure 5**

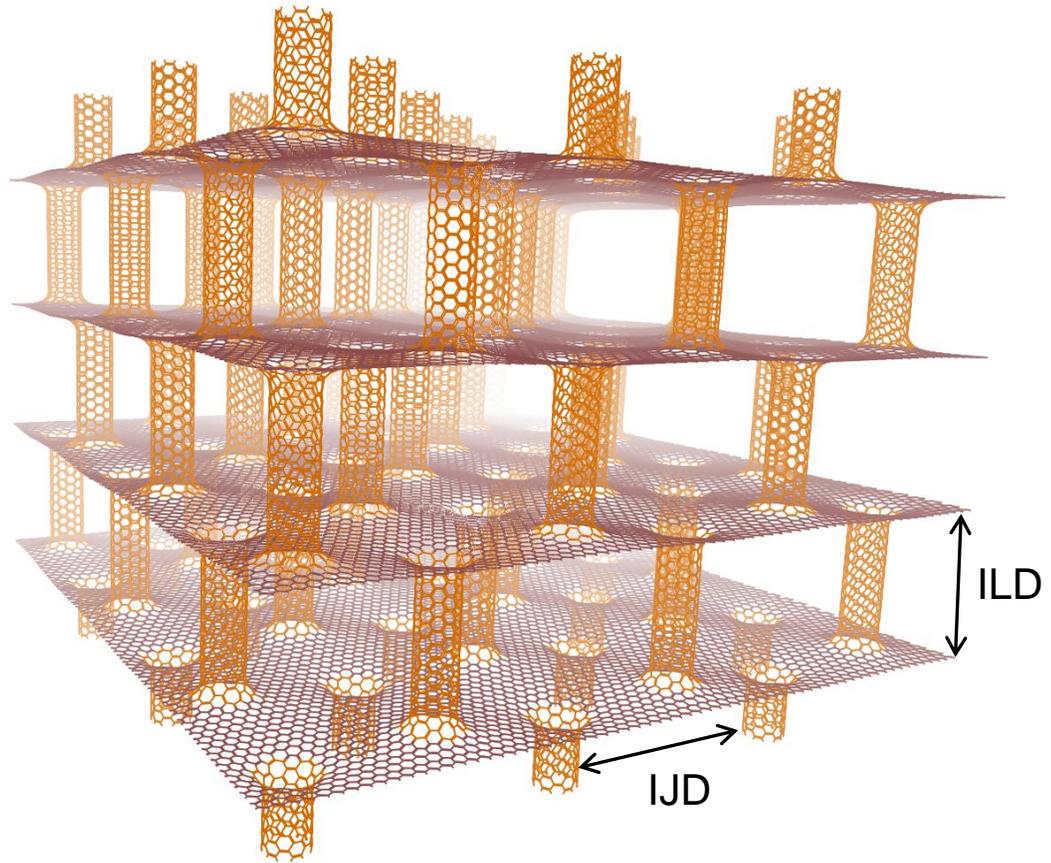

**Figure 6**